# Computational Tools in EN-MME: Implicit and Explicit Finite-Element Simulations

*F. Carra, G. Coladonato, A. Piccini, J. Swieszek, N. Vejnovic.*
CERN, Geneva, Switzerland

**Abstract**
This paper recalls the principles of the finite-element methods (FEM) theory and declines its application in the EN-MME group, for the numerical modelling and study of particle accelerator equipment. Implicit and explicit methods are compared, and practical examples of their use are given.

**Keywords**
FEM, implicit, explicit, simulations.

## 1  Introduction to the Finite-Element Method

For several decades, computer-aided techniques have been widely adopted in both industrial engineering and research environments. Typical examples include Computer-Aided Design (CAD), Computer-Aided Engineering (CAE), and Computer-Aided Manufacturing (CAM). In the scope of this paper, the focus is placed on CAE tools, and in particular on their application to the analysis and design of particle accelerator equipment.

Computer-Aided Engineering (CAE) tools play a central role throughout the design cycle of complex engineering systems. Starting from the initial design and analysis phase, CAE enables extensive virtual prototyping activities, allowing the evaluation of multiple design options at a relatively low cost, compared to physical prototyping and final product testing. As the development process progresses, the cost and effort associated with experimental validation and full-scale testing increase significantly. For this reason, investing time and resources in early-stage numerical analysis is widely recognized as an effective strategy to reduce overall development time, cost, and technical risk.

Within the CAE framework, several complementary simulation domains are commonly employed, depending on the physical phenomena under investigation. These include Finite-Element Method (FEM) for structural, thermal, and electromagnetic studies, Computational Fluid Dynamics (CFD) for fluid flow and heat transfer, and Multibody Dynamics (MBD) for the analysis of mechanical systems with interacting components. More advanced approaches may also involve Multidisciplinary Design Optimization (MDO), where multiple physical domains are coupled to optimize system-level performance. In the context of particle accelerator equipment, FEA-based simulations represent a key component of CAE activities, particularly for assessing structural integrity and thermo-mechanical behaviour under operational and accidental load conditions.

### 1.1  FEM theory: in a nutshell

Let us start with a brief introduction on the FEM basic principles. The main concept of FEM is that the displacements of all the points in a continuum under the action of external forces depends on the displacements of discrete points known as nodes. This dependence is regulated by interpolating functions known as shape functions. To study a body with FEM, we must thus discretize the continuum in a finite number of elements, each one featuring a number of nodes which depends on the type of element chosen. An example of discretization of a common object is given in Fig. 1.

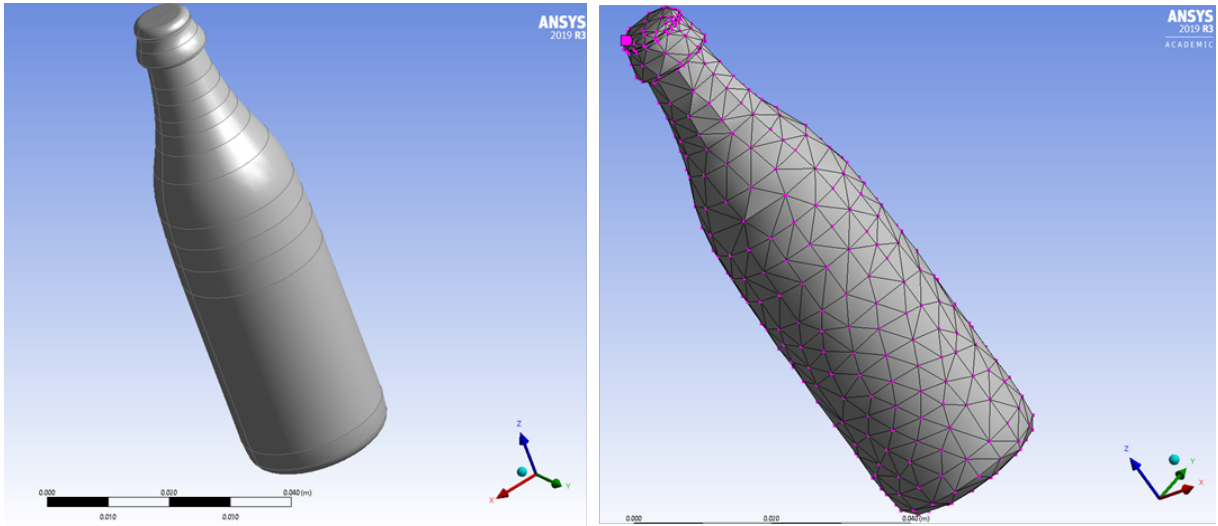

**Fig. 1:** Discretization of a bottle with tetrahedral elements.

The shape functions depend on the type of elements chosen to discretize the physical system. The finite elements, in turn, are dependent on the model adopted to study the system:

- Line elements are selected to model 1D structures like beams, rods or pipes.
- Surface elements are used to model large and thin surfaces like shells, plates.
- Solid elements are used to model three-dimensional bodies.

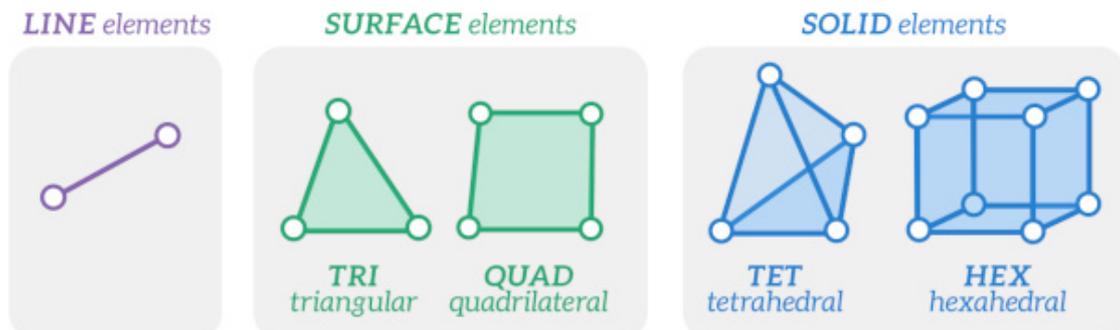

**Fig. 2:** Examples of finite elements.

Within FEM, the primary unknowns of the problem are the nodal degrees of freedom, typically expressed in terms of nodal displacements. Once the system of equations has been assembled, the solution is obtained by solving a linear (or non-linear) system of the form:

$$s = [K]^{-1}F, \qquad (1)$$

where $[K]$ is the global stiffness matrix, while $F$ represents the external load vector. The vector $s$ collects the nodal displacement components for all degrees of freedom of the discretized domain. After the nodal solution has been obtained, the displacement field within each finite element is reconstructed through the use of the shape functions $[N]$. As explained, these functions interpolate the nodal displacements and allow the displacement vector $u$ to be evaluated at any point inside the element as:

$$u = [N]s . \qquad (2)$$

From the displacement field, the strain components are computed by applying the compatibility equations, leading to:

$$\varepsilon = [\partial]u = [\partial][N]s \ . \tag{3}$$

The stress state is then derived from the strain field through the material constitutive law, commonly expressed in matrix form as:

$$\sigma = [D]\varepsilon \ , \tag{4}$$

where $[D]$ is the material stiffness matrix, for instance corresponding to Hooke's law in the case of linear elastic behaviour. The combination of shape functions, compatibility equations, and constitutive relations enables the evaluation of displacement, strain, and stress fields throughout the entire computational domain, and not only at the nodal locations. This capability represents one of the key strengths of the FEM approach and forms the theoretical basis for its widespread use in the structural and thermo-mechanical analysis of complex engineering systems.

## 2 Finite-Element Solvers: Implicit vs. Explicit

In transient finite-element analyses, two main classes of time-integration solvers are commonly employed: implicit and explicit formulations. Explicit solvers compute the unknown quantities (displacements, velocities, and accelerations) at a time instant $t + \Delta t$ by directly evaluating the equilibrium equations at the current time $t$. This approach leads to uncoupled equations and avoids the need for global matrix inversion, typically relying on lumped mass matrices and simple matrix–vector operations. As a result, explicit solvers are computationally efficient per time step and particularly robust in the presence of severe nonlinearities, large deformations, contact, or material failure. However, their stability is conditional, requiring very small timesteps, dictated by the smallest element size and wave propagation speed in the model.

Implicit solvers, on the other hand, determine the solution at $t + \Delta t$ by enforcing equilibrium simultaneously at the current and future time steps, which results in a system of coupled equations that must be solved iteratively, often involving matrix factorization or inversion. This formulation is unconditionally stable with respect to the time step size, allowing the use of relatively large time increments and making implicit solvers well suited for quasi-static problems and slow transient phenomena. The main drawbacks are the higher computational cost per time step and the potential for convergence issues in highly nonlinear regimes. Consequently, implicit and explicit solvers are complementary tools, with explicit formulations generally preferred for fast transients and strongly nonlinear events, and implicit approaches favoured for static or slowly evolving problems.

At CERN, several loading scenarios can be simulated with implicit FE codes, see Fig 3.

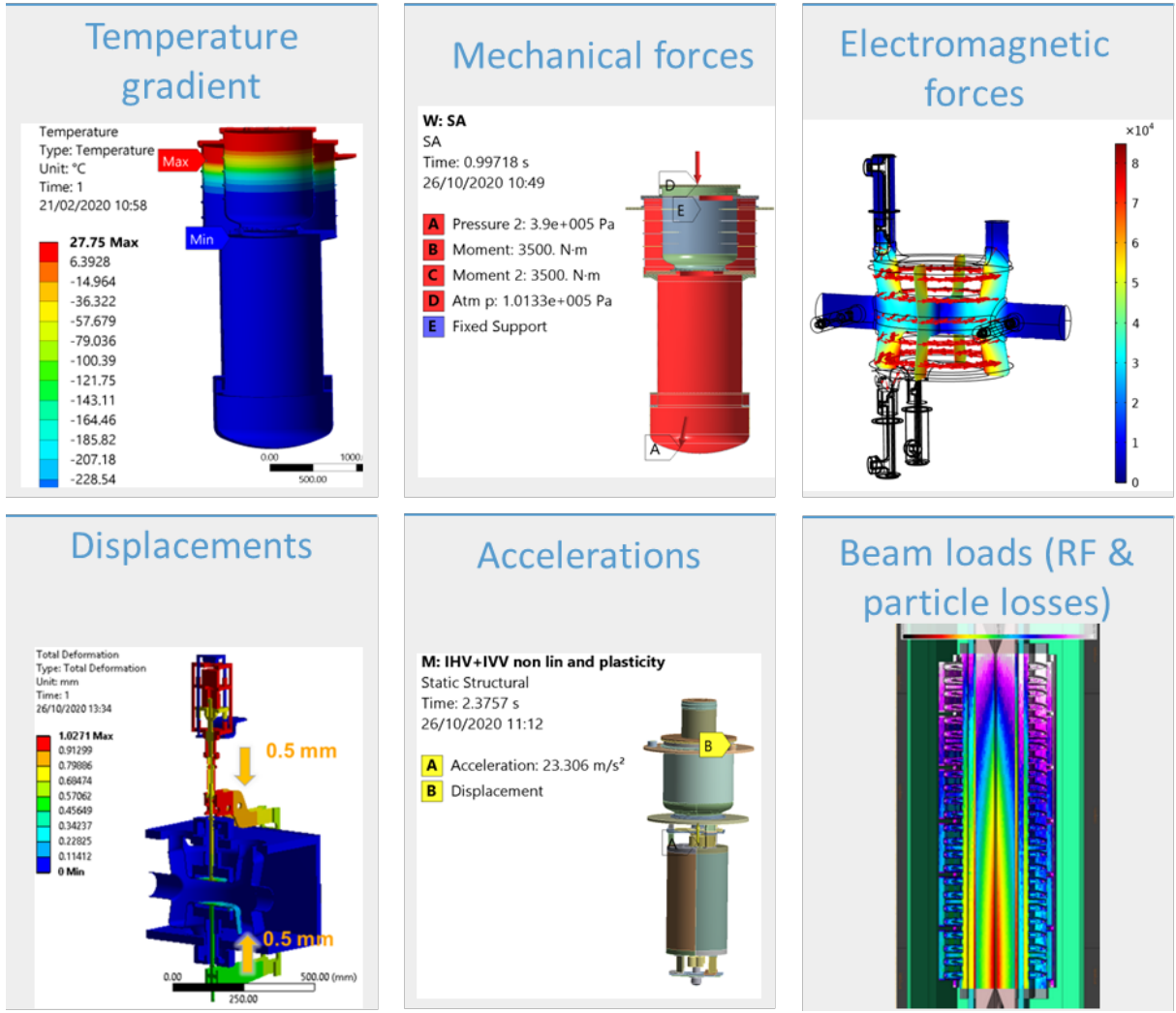

**Fig. 3:** Load cases and engineering domains usually studied with implicit FE codes in EN-MME.

On the other hand, we resort to explicit codes in a number of limited applications, notably when the investigated problem is characterized by high nonlinearities. Examples of the use of implicit and explicit solvers will be given in the next chapters.

## 3  From the real system to its numerical model

Before moving to examples of use of implicit and explicit codes at CERN, EN-MME, we want to describe an important preliminary phase of the engineering study. Before performing any numerical simulation, a fundamental step in the design of a mechanical component is the clear identification and formalization of all relevant loading scenarios. This requires a thorough understanding of how the component operates throughout its entire life cycle, including nominal operating conditions, transitions between different operating states, and non-operational phases such as transport, installation, handling, and maintenance. Particular attention must be paid to the nature of these transitions, distinguishing between slow and fast transients and assessing whether dynamic effects may become significant. In addition, anticipated testing procedures prior to commissioning, the expected number of load cycles, and the influence of environmental factors such as temperature, radiation, humidity, or chemical effects must be considered, as they can significantly affect the mechanical behaviour of the component.

All these aspects should be consolidated into a comprehensive set of load cases, typically documented in a formal specification or *cahier des charges*. In complex systems, such as particle accelerator equipment, this process may lead to the definition of a large number of operational, exceptional, and testing load cases. However, it is often possible to rationalize this set by identifying a limited number of critical scenarios that govern the mechanical design. Numerical simulations can then be focused on these most demanding load cases, ensuring structural integrity and compliance with design requirements while keeping the overall analysis effort manageable. In this way, a specification involving tens of load cases, can often be addressed by performing only few simulations (ideally, and typically, 1-2 operational / exceptional loading conditions, and 1-2 testing scenarios).

After downselecting the number of scenarios to simulate, a second important step is to create a numerical model of the real system. The model needs to be accurate enough to reproduce the expected system behaviour, but not exaggeratedly detailed, as this increases the computational time and size of the solution files. Model simplification can follow a set of common guidelines. For example, geometrical details that do not significantly influence the global response of the structure should be removed in order to reduce model complexity and computational cost (Fig. 4). Features such as screws, bolts, and welds are often not explicitly represented in the global FE model, but instead assessed separately by post-processing internal forces extracted from the analysis or by means of dedicated hand calculations. Local geometrical details such as chamfers and fillet radii can be addressed through submodelling techniques when required. While model simplification is generally straightforward, the definition of loads and boundary conditions represents the most critical step of the entire simulation process. These must reflect the real working conditions as accurately as possible, even though pragmatic compromises are sometimes unavoidable to ensure numerical robustness (for example, simplified representations of nonlinear contacts). Given the inherent uncertainties associated with modelling assumptions, appropriate safety factors should be applied, effectively accounting for a "factor of ignorance". When approximations are introduced, they should always be made on the conservative side. A practical and robust strategy consists in starting from a simple model and progressively increasing its level of detail only where justified by the analysis objectives. In a nutshell: *start simple, complexify later!*

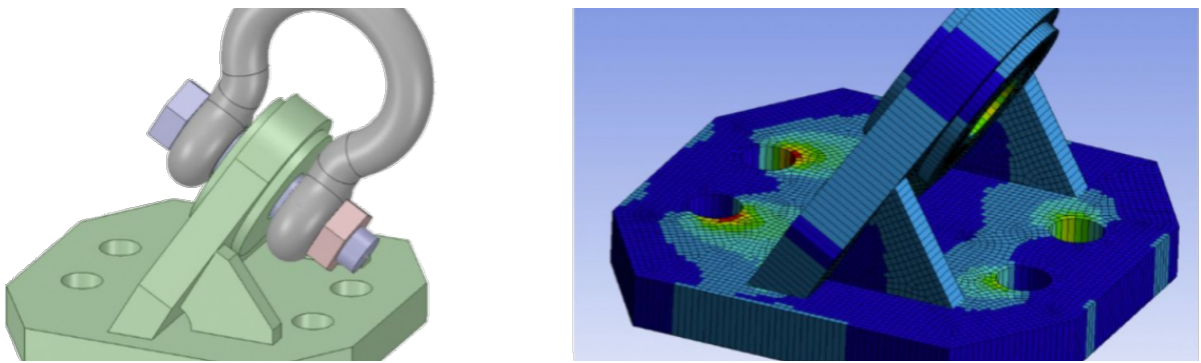

**Fig. 4:** Example of numerical modelling of a lifting element.

## 4    Examples of implicit simulations

As discussed, at CERN EN-MME, implicit FEM are usually adopted to solve problems involving a complex geometry, quasi-static regimes and non-linearities excluding materials' changes of phase. For this type of problems, the tool of preference is ANSYS Workbench [1], which features a powerful geometry generator, as well as a stable solver and a highly detailed post-processor.

On example of project treated with this technique was the design and analysis of the Facility for Reception of Superconducting Cables (FRESCA2) [2]. The facility features a big-sized (about 4 m high, 2 m wide) double-nested cryostat for the measurement of superconducting samples (Fig. 5). The design of the system required a combination of thermal and mechanical simulations, to optimize its cooling efficiency without compromising the structural resistance of the system.

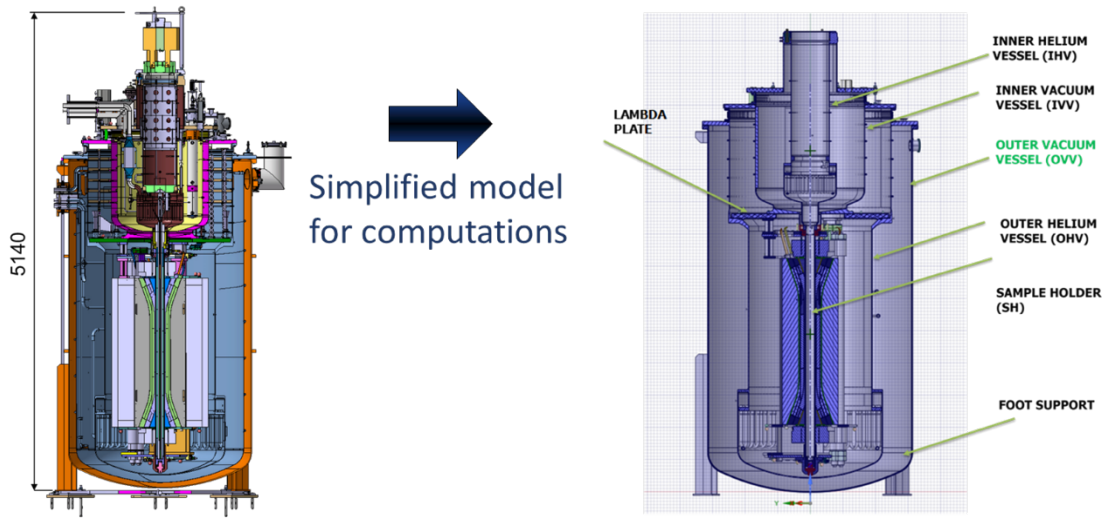

**Fig. 5:** FRESCA2, from detailed to simplified model.

The following paragraphs give a description of the study performed to design the outer helium vessel (OHV) by means of ANSYS Workbench.

### 4.1 Definition of the "Cahier des charges"

The specification of the FRESCA2 OHV included ten operational load cases, and three testing scenarios, for a total of 13 load cases (Fig. 6).

**3 OUTER HELIUM VESSEL (OHV)**

Nominal operation load cases
  NLC1 - Transport
  NLC2 - Installation in the pit
  NLC3 - Assembly
  NLC4 - Vacuum pumping
  NLC5 - Pressurized
  NLC6 - Cold
  NLC7 - Powering
  NLC8 - Quench
  NLC9 - Vacuum loss
  NLC10 - Purge with vacuum loss

Testing load case
  TLC1 - Leak test during fabrication
  TLC2 - Pressure test during fabrication
  TLC3 - Pressure test in place

**Table 2 - Applicable load cases for outer helium vessel**

|  | NLC1 | NLC2 | NLC3 | NLC4 | NLC5 | NLC6 | NLC7 | NLC8 | NLC9 | NLC10 | TLC1 | TLC2 | TLC3 |
|---|---|---|---|---|---|---|---|---|---|---|---|---|---|
| Self-weight (1) | A | B | B | B | B | B | B | B | B | B | C | C | B |
| Temperature (2) | A | A | A | A | B | B | B | B | A | A | A | A | A |
| Internal pressure (3) | / | / | / | A | B | B | B | C | C | / | / | D | E |
| External pressure (4) | / | / | / | / | / | / | / | B | B | A | A | / | / |
| Magnet + IC weight | / | / | X | X | X | X | X | X | X | X | / | / | X |
| Torque | / | / | / | / | / | X | X | X | / | / | / | / | / |

(1) A = self-weight supported by handling points; B = self-weight supported by top flange;
    C = self-weight on manufacturing supports
(2) A = 300 K; B = 4.5 – 300 K thermal gradient
(3) A = Atmospheric pressure; B = 1.3 bar (absolute); C = PS (3.9 bar absolute);
    D = Hydraulic Test pressure (1.43 x PS); E = Pneumatic test pressure (1.25 x PS)
(4) A = Atmospheric pressure; B = 1.5 bar (absolute)

**Fig. 6:** FRESCA2 OHV – Cahier des charges.

The analysis of the cahier des charges helped reducing the 13 load cases to only two design cases: Quench during operation (during which a 3.9 bara internal pressure acts on the shell, on top of a thermal gradient, and an electromagnetic torque), and Vacuum loss during purging (where the system must sustain an external pressure of 1.5 bara). It is not possible to further downselect between these two scenarios because, a priori, it is not straightforward to determine which load case is worst between these two. Their failure mode are, in fact, different: for the first one, the expected failure scenario is by plastic deformation, whereas for the second load case, it is easy to expect buckling to be the limiting failure mechanism.

## 4.2    Quench during operation

The thermal field generated during the quench event was first computed through a dedicated thermal analysis and subsequently imported into the structural model, allowing a decoupled thermo-mechanical approach (Fig. 7). In the preliminary design phase, simplified modelling assumptions were adopted, such as the use of shell elements and linear elastic material behaviour, while more advanced nonlinear effects were introduced at a later stage. Structural verification was performed in accordance with EN 13445-3 Annex B (Direct Route) [3], which requires the maximum principal strain to remain below 5%. The numerical results showed that the maximum strain levels remained well within the allowable limit (Fig. 8). The accuracy of the analysis was assessed through mesh convergence studies and local submodelling, ensuring the robustness of the obtained results.

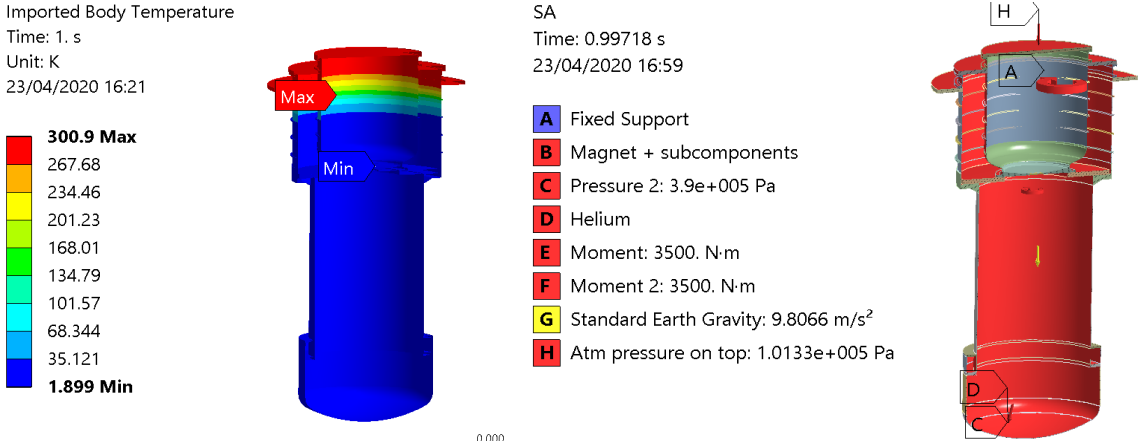

**Fig. 7:** FRESCA2 OHV, loads applied in the structural simulation.

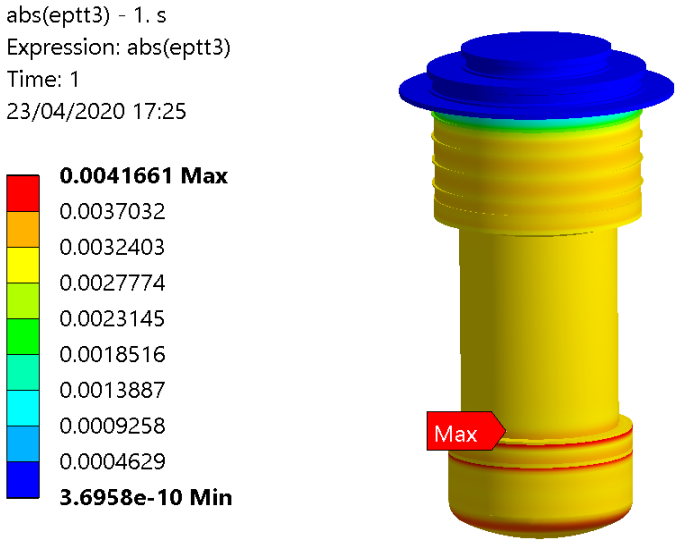

**Fig. 8:** FRESCA2 OHV, highest absolute principal strain.

## 4.3 Vacuum loss during purging

In this case, one critical failure case is represented by buckling. There are several ways, of increasing difficulty, to simulate buckling (Fig. 9). The easiest is a simple eigenvalue / eigenvector simulation, also known as elastic bifurcation analysis, or linear buckling analysis. It is typically easier to start with this analysis in the preliminary design phase, aiming at relatively high "safety factors" (> 10-15), as this method can be pretty inaccurate to predict the failure of complex geometries, different from simple spherical, conical or cylindrical perfect shapes. In the last stage of the equipment validation, a complete nonlinear buckling analysis must be performed, activating nonlinearities in terms of materials, large deformations and initial geometrical errors. These nonlinearities usually significantly decrease the safety factor predicted by the simple linear buckling analysis. In the case of the FRESCA2 OHV, the nonlinear buckling analysis predicted a failure at an external pressure close to 3 bar (sufficiently higher than the specified external pressure acting on the equipment), see Fig. 10.

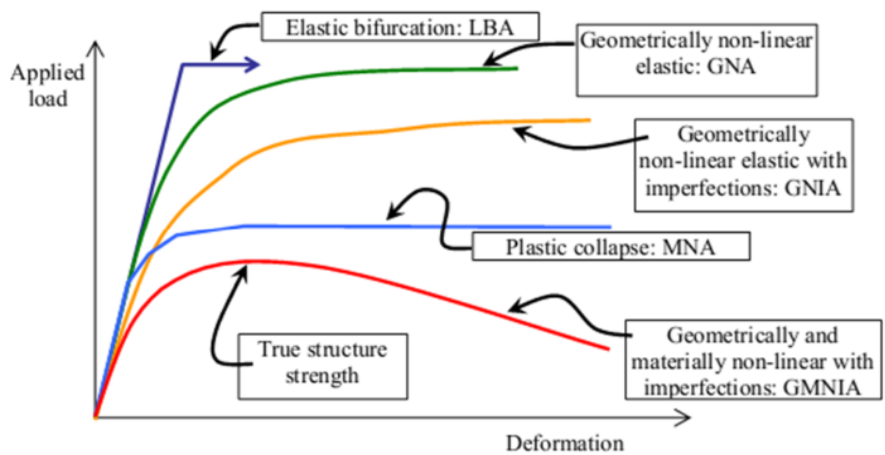

**Fig. 9:** Methods to simulate buckling.

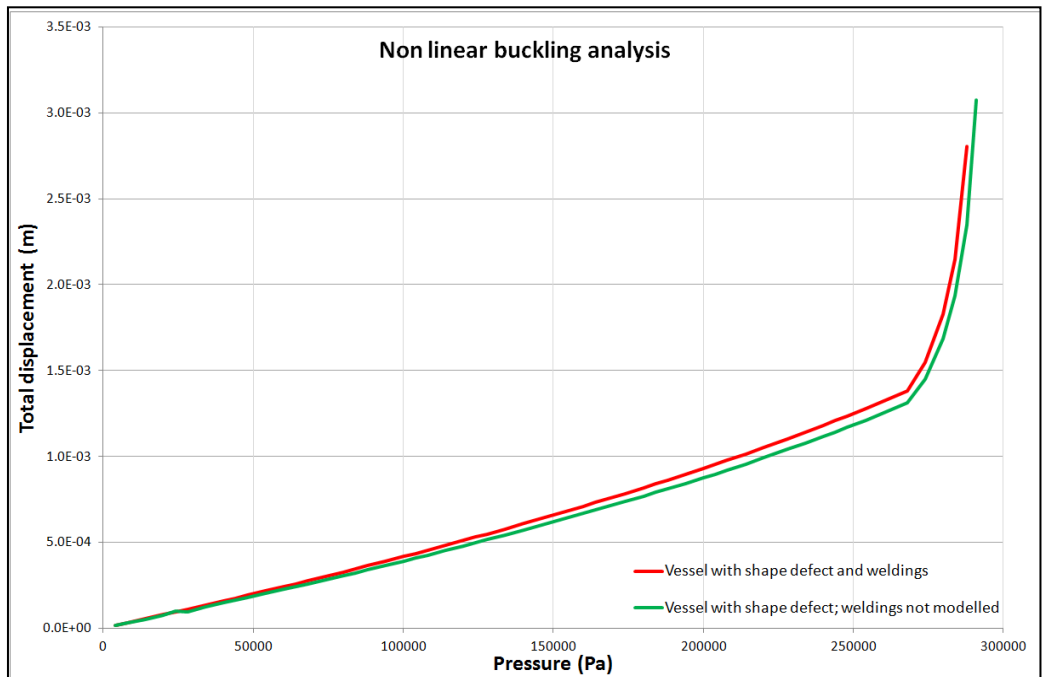

**Fig. 10:** Nonlinear buckling of a shell under external pressure.

# 5   Examples of explicit simulations

We discussed in Chapter 2 about advantages and drawbacks of implicit and explicit solvers. As a reminder, explicit solvers are a powerful tool to reproduce high nonlinearities and material failure, up to catastrophic failures such as fragmentation, explosion, vaporization, etc. This is thanks to the Equation of State (EOS) implemented in an explicit code, which covers the material behaviour at extreme ranges of pressure, density and temperature. The EOS is then coupled with advanced failure models, able to cover the abovementioned phenomena.

However, the main drawback of explicit tools is their conditional stability. In other words, the method is stable only below a critical time step which is calculated as the minimum time step needed for a stress wave to propagate between one element and the adjacent one in the FEM model.

In its simplest form, the critical time step can be expressed as:

$$\Delta t < \frac{L}{c} \approx L\sqrt{\frac{\rho}{E}}, \quad (5)$$

where $L$ is the characteristic element length, $c$ is the elastic wave speed, $\rho$ the material density, and $E$ the Young's modulus. For typical structural materials such as steel and standard mesh sizes, this condition leads to extremely small time increments. As an illustrative example, a model discretized with 1×1×1 mm³ elements in steel results in a critical time step of the order of 0.2 µs. While this does not represent a major drawback for very fast transient phenomena, such as particle beam impacts with characteristic durations of a few microseconds, it becomes prohibitive for slow transient or quasi-static problems. In such cases, the required number of explicit integration steps can easily exceed $10^5$ for low-frequency dynamic responses, or even $10^8$ for quasi-static simulations, making explicit approaches computationally inefficient. This fundamental limitation explains why explicit solvers are best suited for short-duration, high-rate events, whereas implicit formulations are generally preferred for slow transient and static analyses.

The preferred adoption of an implicit or explicit solver can be sketched as in Fig. 11.

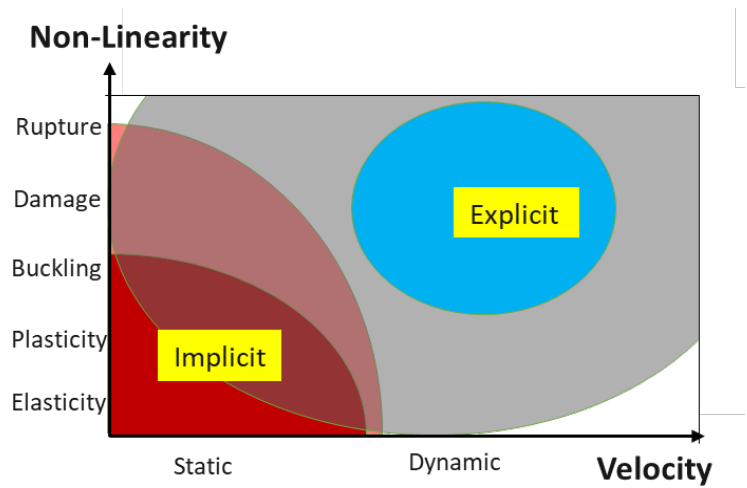

**Fig. 11:** Graphical summary of the preferred solvers for different engineering problems.

Very fast CERN-characteristic events, such as the impact of the particle beam on a target, can only be simulated with an explicit code. There are cases, however, where the choice of adopting one solver or the other is less straightforward. For example, fabrication processes like deep drawing, bending, forming, generate very significant nonlinearities, up to fracture, but they are somehow slow (quasi-static, or slow-transient), as they can last few seconds or even minutes. This duration, compared to the maximum time step calculated before with (5), is extremely long for a simulation with an explicit code. On the other hand, the nonlinearity of these fabrication processes are often too high to be correctly reproduced with an implicit code. We are thus in an uncomfortable regime, where the choice of one solver of the other is not an easy decision.

At CERN EN-MME, the preference tool for the simulation of fabrication processes is the explicit code LS-Dyna [4]. Using this method, we need to compensate the issue of the critical time step being much smaller than the process duration. With reference to Eq.(5), and considering that the computational time depends on the number *n* of time steps needed to simulate the process duration $t_{tot}$, a few techniques need to be adopted:

1. Increase of the mesh size *L*, at the risk of losing in accuracy at the locations with peak stresses.
2. Artificial increase of the density $\rho$, also known as mass scaling, adding mass to the model in specific locations.
3. Artificial decrease of the process duration $t_{tot}$, also known as time scaling, increasing the speed of the numerical process with respect to the real one.

Both *2.* and *3.* present the significant risk of introducing artificial dynamic effects to the simulation. Such risk must be monitored and ruled out with specific checks: for more on this topic, we refer to [5].

In the next two chapters, we report one example of fabrication process treated with LS-Dyna, and one case of quasi-instantaneous event, simulated with Autodyn.

### 5.1 Fabrication processes: hydroforming

Hydroforming is a manufacturing process widely used for the production of accelerating cavities, in which a precursor tube is shaped into the final geometry by the combined action of internal hydraulic pressure and axial compression applied through dedicated tools and moulds (Fig. 12). The process is often carried out in multiple stages and may include intermediate thermal treatments such as annealing. Due to the large number of interacting parameters – including pressure, applied forces, tool geometry, strain rate, temperature, and friction – the optimization of the hydroforming process cannot rely solely on experimental trial-and-error approaches, which would be time-consuming and costly. In this context, explicit finite-element simulations, using LS-DYNA, provide an effective tool to reduce the parameter space and to support comparative and optimization studies during process development.

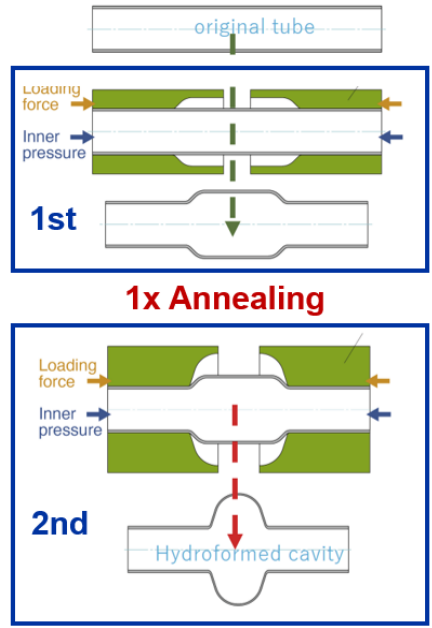

**Fig. 12:** Elliptical cavity hydroforming sequence.

In this case, the simulation helped in defining the optimal process parameters, ruling out the risk of buckling that occurs in both hydroforming steps, see Fig. 13.

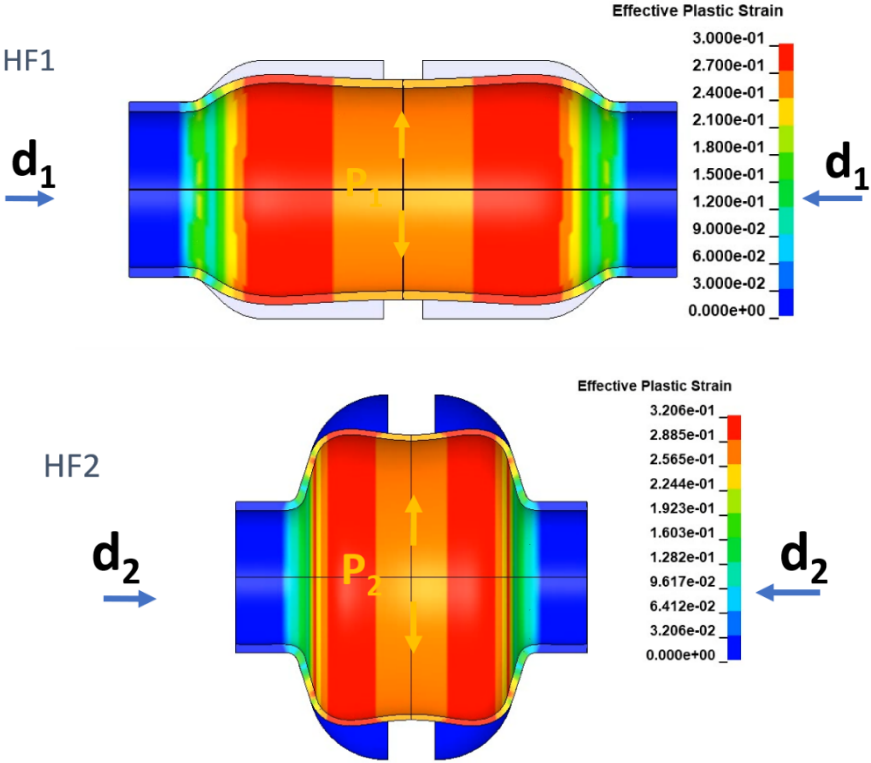

**Fig. 13:** LS-Dyna simulation of copper elliptical cavity hydroforming.

Thanks to the iteration between experimental and numerical techniques at EN-MME, in collaboration with KEK, the fabrication problems occurring in the development phase of the process were successfully overcome, leading to the production of four seamless cavities in a few hours.

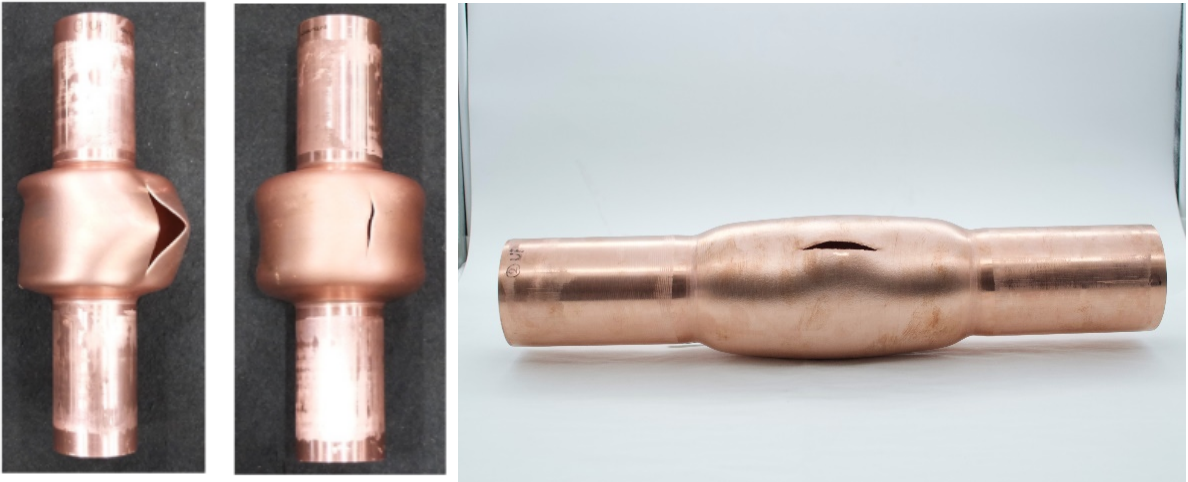

**Fig. 14:** Fracture of cavity samples in the initial phase of the hydroforming R&D.

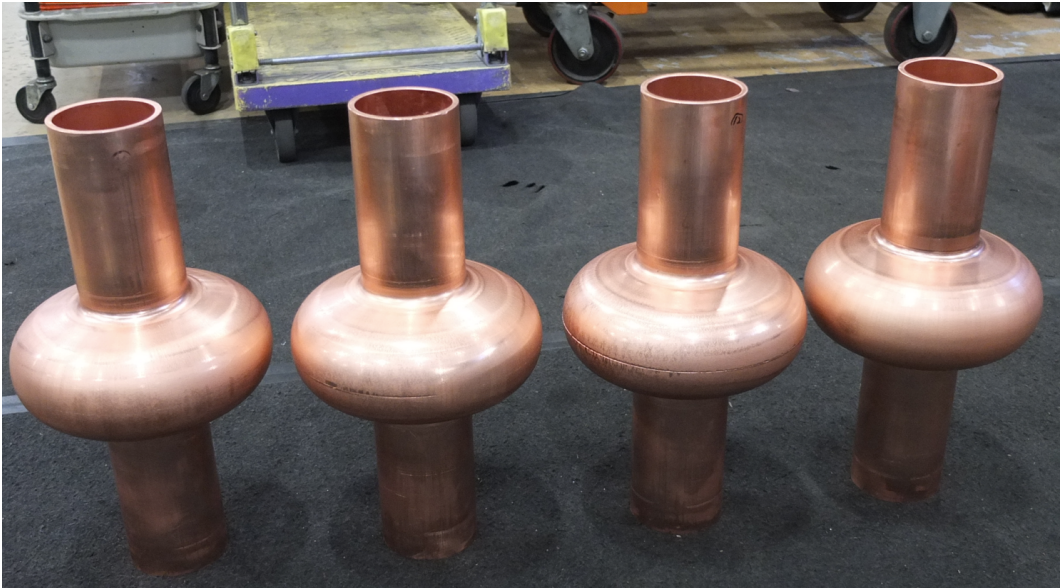
**Fig. 15:** Four seamless cavities produced at KEK (M. Yamanaka, A. Yamamoto) after optimization of the process.

## 5.2  Response of structures under particle beam impact

When a particle beam impacts an accelerator component, its kinetic energy is rapidly transferred to the material in the form of heat. Due to the extremely short duration of the impact – typically in the nanosecond to microsecond range – this process is commonly referred to as *quasi-instantaneous heating*. In this regime, the characteristic deposition time $t_d$ is much smaller than both the thermal and mechanical time constants of the structure, i.e.:

$$t_d \ll \tau_{\text{th}}, \tau_{\text{mech}}$$

The thermal time constant can be estimated as:

$$\tau_{\text{th}} = \frac{L^2}{a}, a = \frac{k}{\rho c_p},$$

where $L$ is a characteristic length, $a$ the thermal diffusivity, $k$ the thermal conductivity, $\rho$ the density, and $c_p$ the specific heat. The mechanical time constant is given by

$$\tau_{\text{mech}} = \frac{L_c}{c}$$

with $L_c$ the characteristic dimension of the system and $c$ the elastic wave speed.

Such events are characterized by very fast transients and strong nonlinearities, potentially involving melting, vaporization, and fragmentation, which makes them a typical application domain for explicit numerical solvers. At CERN, particle beam impact scenarios are commonly analysed using the explicit hydrocode Autodyn, including advanced material models and dedicated equations of state [6].

The study is divided into two phases. First, the thermal energy deposition on the target must be evaluated with a Monte-Carlo interaction and transport code, such as GEANT4 [7], MARS [8], or FLUKA [9]. Then, the 3D energy density map can be used as input for the explicit Autodyn simulation (Fig. 16).

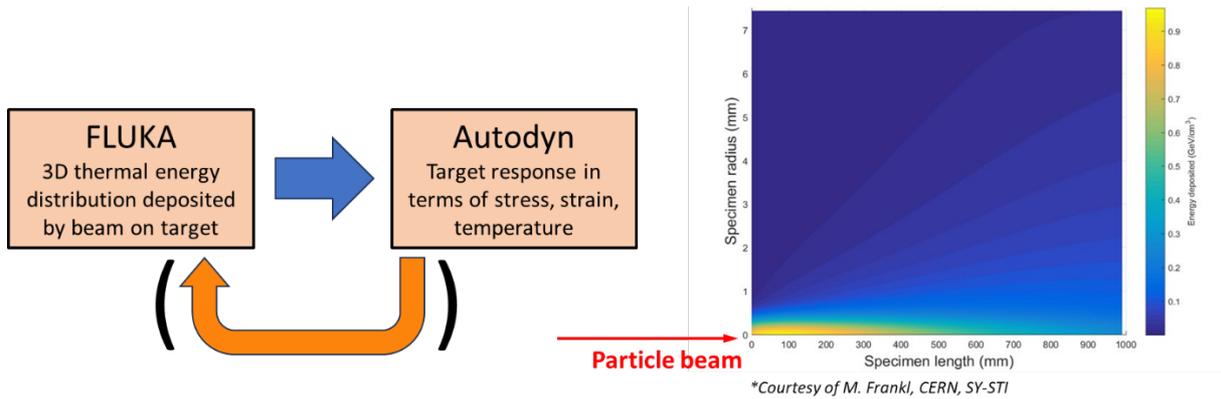

**Fig. 16:** Left, interaction between transport particle code and explicit code. in case of significant density change during the particle beam impact, a coupling between the two codes must be required. Right, example of 440 GeV proton beam impact on a graphite cylinder (FLUKA).

The Autodyn simulation can provide, as results, elastic and plastic stress and strain fields, as well as extreme phenomena that entail melting, vaporization, fragmentation of the target (Fig. 17 and Fig. 18).

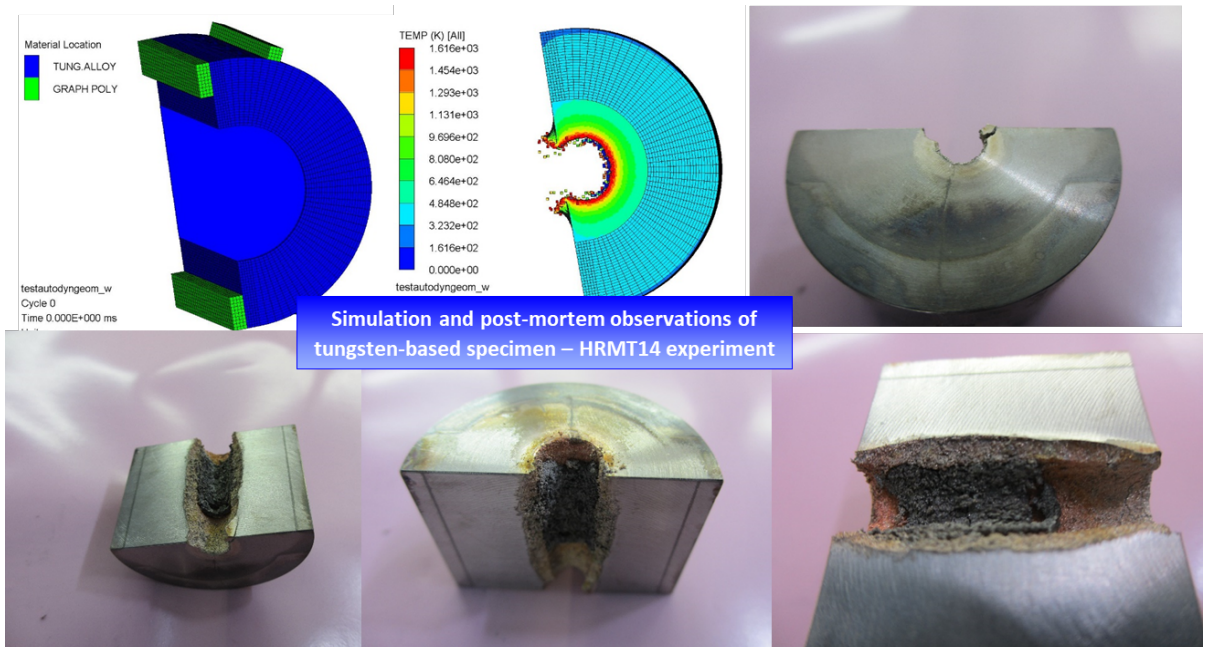

**Fig. 17:** Comparison of Autodyn simulations and experiment at CERN HiRadMat, for the case of a tungsten cylinder impacted by 440 GeV proton beam.

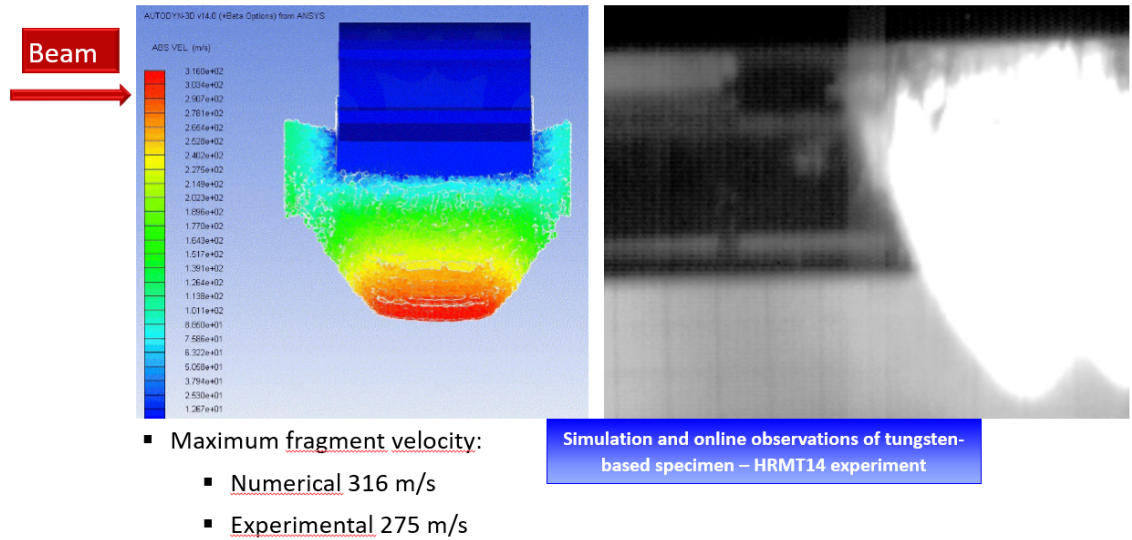

**Fig. 18:** Fragment front propagation speed simulated with Autodyn (left) and captured with high-speed camera at HiRadMat (right).

# 6    Conclusions

This paper has reviewed the basic principles of the Finite-Element Method and their application within the EN-MME group for the engineering analysis of particle accelerator equipment. The role of numerical simulations throughout the design workflow has been discussed, from the definition of load cases and modelling assumptions to the verification of structural integrity under operational, exceptional, and testing scenarios.

Implicit finite-element simulations were first addressed, as they represent the primary numerical tool for most EN-MME applications. Their suitability for quasi-static and slow transient problems involving complex geometries was illustrated through the design and validation of the FRESCA2 Outer Helium Vessel. This example highlighted the importance of a clear cahier des charges, appropriate model simplifications, and progressive refinement of the analysis, including nonlinear material behaviour and buckling verification at later design stages.

Explicit finite-element simulations were then introduced as a complementary approach, required for problems characterized by very fast transients and strong nonlinearities, such as fabrication processes and particle beam impact scenarios. The limitations imposed by conditional stability and critical time step constraints were discussed, together with practical strategies such as mesh optimization, time scaling, and mass scaling. The application of explicit solvers to hydroforming processes and beam-induced thermomechanical loading demonstrated their capability to address extreme regimes involving large plastic deformations, material failure, and phase changes.

Overall, implicit and explicit FEM simulations should be regarded as complementary tools rather than competing approaches. When used with a clear understanding of their respective assumptions and limitations, and in combination with experimental validation and engineering judgement, they constitute an essential asset for the safe design, optimization, and operation of accelerator components at CERN.

**Acknowledgement**

The authors wish to thank all the colleagues that contributed to the examples summarized in this paper.

# References

[1] https://www.ansys.com/.

[2] M. Bajko *et al.*, *Upgrade of the CERN Superconducting Magnet Test Facility*, IEEE Transactions on Applied Superconductivity, vol. 27, no. 4, Jun. 2017, doi: 10.1109/TASC.2016.2635119.

[3] EN 13445-3:2014, *Unfired pressure vessels – Part 3: Design*, Annex B – Design by Analysis: Direct Route.

[4] J. Swieszek, R&D and Numerical Simulations for SRF Fabrication technologies, 5th SRF Workshop 2022, 3-4 February, https://indico.cern.ch/event/1103356/.

[5] J. A. Zukas, *Introduction to Hydrocodes* Elsevier, 2004.

[6] F. Carra, *Numerical Modelling of High Energy Particle Beam Interactions with Matter*, PhD Thesis, Politecnico di Torino / CERN, 2018, https://doi.org/10.5281/zenodo.1410583.

[7] J. Allison *et al.*, *Recent developments in GEANT4*, *Nucl. Instrum. Meth. A* 835 (2016).

[8] N. V. Mokhov *et al.*, *MARS15 Overview*, *Prog. Nucl. Sci. Technol.* 2 (2011) 100–105.

[9] A. Fassò, A. Ferrari, J. Ranft, P. R. Sala, FLUKA: a multi-particle transport code, CERN-2005-10, INFN/TC_05/11, SLAC-R-773 (2005).